# Hybrid Cathode Lithium Battery Discharge Simulation for Implantable Cardioverter Defibrillators Using a Coupled Electro-Thermal Dynamic Model

Mahsa Doosthosseini*, Mahdi Khajeh Talkhoncheh, Jeffrey L. Silberberg, Sandy Weininger, Hamed Ghods

*Abstract*— This paper investigates the impact of implantable cardioverter defibrillator's (ICD's) load on its lithium battery power sources through a coupled electro-thermal dynamic model simulation. ICDs are one of the effective treatments available to significantly improve survival of patients with fatal arrhythmia (abnormal heart rhythm) disorders. Using a lithium battery power source, this life-saving device sends electrical shocks or pulses to regulate the heartbeat. The service life and reliability of an ICD is primarily expressed by its battery's lifespan and performance. In this paper we investigate the terminal voltage, depth of discharge (DOD) and temperature dynamics of the implantable lithium battery with a combined cathode material, namely carbon-monofluoride and silver vanadium oxide (Li/CFx-SVO). Modeling the implantable batteries characteristics is a well-established topic in literature; however, to the best of the author's knowledge, the impact of the high-energy shocks (defibrillation) and low-energy device power supply (housekeeping) on the ICD's battery operation is relatively less-explored. Our analysis reveals that the battery terminal voltage is primarily affected by small but continuous housekeeping discharge current in the range of $\mu A$, rather than intermittent high defibrillation current demand in the range of several amps. The results can be used to improve the device design control and operation, thus extending the service life in patients and reducing the need for invasive replacement surgery.

*Index Terms*— Implantable device power source, Li/CFx-SVO battery, Thermo-electrochemical battery model, Optimal discharge current

## I. INTRODUCTION

This article presents a simulation analysis on the lithium battery dynamics in an implantable cardioverter defibrillator (ICD). Cardioverter defibrillators are medical devices for treating ventricular fibrillation and ventricular tachycardia [1], [2]. Ventricular fibrillation refers to uncoordinated contraction of ventricular muscles which is a potentially lethal arrhythmias (abnormal heart rhythm), and ventricular tachycardia refers to a heart rate over 100 beats per minute, which can cause cardiac arrest and thus cessation of blood flow [3]. In the United States alone, sudden cardiac arrest accounts for more than 450,000 deaths each year [4]. The impact of the Covid-19 pandemic also increased out-of-hospital cardiac arrests by 119% compared with earlier control periods, as determined in an analysis over 10 countries [5]. An ICD continuously tracks a patient's heartbeat and monitors for tachycardia (rapid heartbeat), which can be caused by ventricular fibrillation (quivering of the lower heart chambers). The ICD applies a high-energy shock to the heart if it detects ventricular fibrillation that cannot be treated by cardiac pacing therefore it prevents cardiac arrest.

Fig. (1) illustrates an ICD device's components, and its electrode leads implanted into the heart's right atrium and ventricle through the central veins. A simplified form of an ICD is comprised of a lithium battery, a capacitor to store and discharge energy to deliver defibrillation pulses to the heart, high and low voltage circuits including a transformer and a microprocessor, integrated circuits for measuring electrograms, capturing and storing data, and controlling the delivery of therapy. Also, a header connects the electrode leads to the endocardium (the innermost layer of the tissue that lines the chambers of the heart). All these components are encased in a titanium can. The can volume is within the range of 30 to 40 $cm^3$ [6]. These components work together to provide ICD with its essential features. [7], [8], [9], [10].

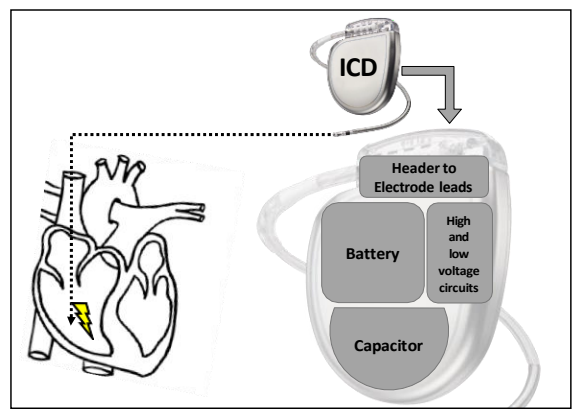

Fig. 1: Implantable Cardioverter Defibrillator (ICD).

The power source of ICDs is high-rate lithium batteries, including lithium manganese oxide (LiMnO2), lithium silver vanadium oxide (Li/SVO), and lithium silver vanadium oxide and carbon monofluoride hybrid (Li/CFx-SVO) [11]. The Li/CFx-SVO battery chemistry is currently being used by a majority of defibrillator manufacturers, due to its high-power density and more predictable end-of-service properties. In this paper we chose to study Li/CFx-SVO batteries for ICDs with the goal of better understanding the impact of ICD's momentary high energy demand and continuous low energy demand on battery dynamics by means of a coupled electro-thermal model of the Li/CFx-SVO battery. This is a well-established line of research for lithium-ion batteries with high capacity applications, such as electric vehicles, but less-explored for lower-capacity applications specifically for

All authors are with the Center of Device and Radiological Health (CDRH) at FDA, Division of Biomedical Physics (DBP), White Oak Campus, silver Spring, MD 20904, USA
*mahsa.doosthosseini@fda.hhs.gov
*Address all correspondence to this author

TABLE I: Batteries used in ICD devices [11], [13], [14], [15]

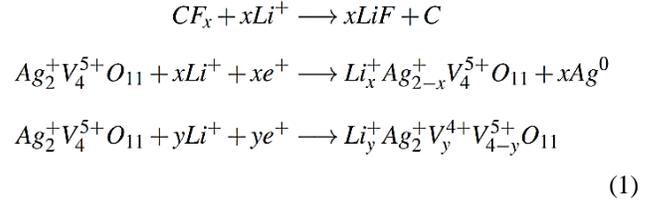

| Battery Type | Theoretical gravimetric capacity of cathode material | Energy density of the cell |
|---|---|---|
| Li/CFx−SVO | 420-717 ($mAh/g$) | 1000 ($mWh/cm^3$) |
| Li/SVO | 272 ($mAh/g$) | 522 ($mWh/cm^3$) |
| LiMnO2 | 192 ($mAh/g$) | 230 ($mwh/g$) |

implants.

Despite the successful attempts for modeling, parameterization, and utilization of Li/CFx-SVO batteries for ICDs, one major challenged still exists, evident from references including Manolis et al. [12]: manufacturers of ICDs promise a 5 to 8 years projected longevity; however, real-life data suggest that not all ICDs reach this lifespan. They studied a population of 685 patients with ICDs and found that 8% of the devices demonstrate premature battery depletion by 3 years. Simulation of the ICD battery performance and longevity over its lifetime with an average number of defibrillation pulse generations and a constant low current to power the device's electronics can improve our knowledge about possible scenarios of battery depletion in ICDs.

In Section II we study the voltage characteristics of the Li/SVO-CFx battery. The open-circuit voltage versus the depth of discharge (DOD) of the battery is also obtained in this section. Section III presents a simulation model for the ICD's battery using a coupled thermo-electrical dynamic model of lithium batteries. Section IV simulates the Li/SVO-CFx model for an average ICD load and also varying defibrillation and device supply currents to study their impact on battery dynamics and lifespan. Finally, Section V summarizes the paper's conclusions.

## II. MATERIALS AND METHODS

Many battery-powered, active implantable medical devices (AIMDs) such as cardiac pacemakers, neurostimulators, drug pumps, and cardiac defibrillators rely on lithium batteries for power. Lithium batteries in AIMD are of interest in this paper due to their high safety and reliability requirements, specifically for ICD applications. Table (I) represents three common lithium battery chemistries for ICDs as discussed in [11].

In Li/CFx-SVO batteries, the hybrid cathode design allows the device to provide shock therapy by providing a higher amount of current. Moreover, Li/CFX-SVO has a stepped voltage profile, due to its multiple oxidation states of silver and vanadium. This can be used as a warning signal when a device needs to be replaced at the end of its service life. [13], [14]. The $Ag_2V_4O_{11}$ is an ideal choice of silver vanadium oxide's stoichiometries developed for ICDs cathode material of ICDs [16], [17]. Hybrid cathodes are developed to improve the power density compared to the Li/CFx system. The discharge behavior of Li/CFx–SVO hybrid cells is described by Gomadam et. al. [18]. Three main chemical reactions occur in the cathode side when the battery discharges:

$$CF_x + xLi^+ \longrightarrow xLiF + C$$
$$Ag_2^+V_4^{5+}O_{11} + xLi^+ + xe^+ \longrightarrow Li_x^+Ag_{2-x}^+V_4^{5+}O_{11} + xAg^0$$
$$Ag_2^+V_4^{5+}O_{11} + yLi^+ + ye^+ \longrightarrow Li_y^+Ag_2^+V_y^{4+}V_{4-y}^{5+}O_{11}$$
(1)

These reduction reactions are irreversible. Lithium insertion in carbon-monofluoride silver vanadium oxide hybrid cathode is accompanied by reduction of CFx, Ag and V. Li metal in the anode oxidizes to produce electrons that travel to the cathode, where they reduce CFx, Ag, and V. The $Li^+$ will then be intercalated into the cathode layers. The intercalated $Li^+$ will defluorinate the CFx and nucleate the formation of xLiF particles and conductive carbon where x is the moles of $Li^+$ intercalated into the material [14]. During discharge, in multistage reduction reactions V5+ reduces to V4+, and $Ag^+$ reduces to $Ag^0$ which make a multi-plateau cell potential profile as shown in Fig. 2 [19]. This characteristic cell potential profile is significant for use in ICD applications as it indicates when battery replacement will be needed [14]. In the next section we apply this voltage curve to an electro-thermal model to study dynamics of the Li-/CFx-SVO battery under ICD's operational conditions.

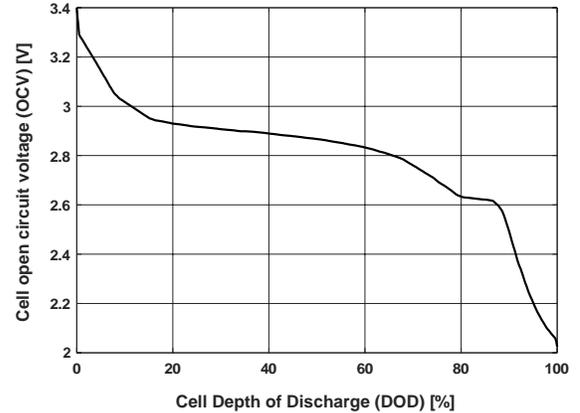

Fig. 2: Open circuit voltage curve as a function of the battery's depth of discharge for a 2:1 hybrid CFx/SVO cathode [19].

## III. MODEL FORMULATION OF THE LITHIUM BATTERY

Consider the following third-order model that represents the dynamics of the lithium battery in a lumped parameter equivalent circuit model (ECM) coupled with a temperature dynamic model:

$$\begin{aligned}
\dot{x}_1 &= \frac{1}{Q}I \\
\dot{x}_2 &= \frac{x_2}{R_1 C_1} + I \\
\dot{x}_3 &= \frac{hA}{mC_p}(T_{amb} - x_3) + \frac{R_2}{mC_p}I^2 \\
y &= OCV(x_1) + \frac{x_2}{C_1} + R_2 I
\end{aligned} \quad (2)$$

The ECM models benefit from simplicity and computationally inexpensive implementation for online state and parameter estimation algorithms for lithium battery management systems and for predicting the battery's future performance [20]. However, to the best of the author's knowledge, this model has not applied for implantable devices applications.

The choice of including multiple RC circuits in a high-order RC network helps the model to be more accurate, but it also increases the number of unknown parameters and states. In this study, we only considered a single RC circuit for our model. Fig. (3) shows the battery model connected to the ICD. The variable I is the discharge current, Q is the capacity of the cell. C1 is the polarization capacitance, R1 is the polarization resistance and R2 is the internal ohmic resistance. The $C_d$ represents the ICD defibrillation capacitor which is charged by the battery for shock therapy. The RC pair represents the battery diffusion dynamics. In this model, three state variables $x_1(t)$, $x_2(t)$, and $x_3(t)$ denote the depth of discharge (DOD), transient voltage and internal battery temperature, respectively. The nonlinear dynamics of the lithium battery cell temperature $x_3$ is coupled to the ECM model through the temperature dynamics model developed by Bernardi *et al.* [21]. Coupling the ECM model with the thermal model for lithium batteries is also used in the literature for battery parameter identifiability analysis and parameter estimation [22].

The $x_3$ dynamics captures three key phenomena. First, the term $mC_p \frac{dT}{dt}$ represents the battery cell's thermal energy storage ability, where $T$, $m$, and $C_p$ denote the cell's temperature, mass, and specific heat capacity, respectively. Second, the term $hA(T_{amb} - x_3)$ represents convective heat transfer between the battery cell and the surrounding, where $h$, $A$, and $T_{amb}$ denote the convection heat transfer coefficient, convection area, and the patient's body temperature, respectively. Third, irreversible heat generation is represented by $R_2 I^2$, where $R_2$ is the cell's ohmic resistance. Nominal parameters for this model, corresponding to a commercial ICD sized $li/SVO - CFx$ battery cell, are taken from Gomadam *et al.* [18] and shown in Table (II) below.

The last term of Eq. (2) is the battery terminal voltage ($y$) which is the output equation of the model. $y$ is the summation of the cell's open circuit voltage as a function of DOD, the voltage drops across the $R_1 C_1$ pair and the voltage drop across the internal resistance ($R_2$).

We use this model to simulate the ICD's lithium battery discharge behavior over the course of 5 to 7 years to model the impact of occasional high-energy defibrillation pulses and a continuous low-energy housekeeping current on the battery performance. The open circuit voltage of the battery as a function of the depth of discharge $OCV(x_1)$ in the output equation is a unique curve for each battery chemistry that represents the electrochemical characteristics of the reduction reactions during discharge (or charge). The OCV-DOD curve of Li/SVO-CFx batteries is presented in Section III, and we use it in our model to study the ICD load impact on the battery performance and longevity.

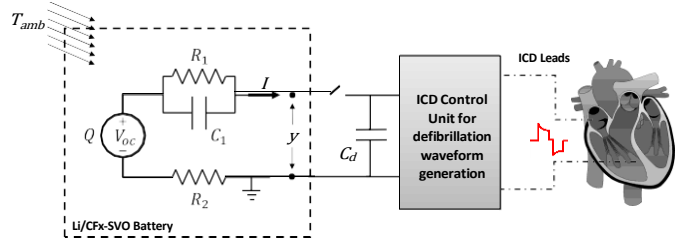

Fig. 3: The thermo-electrical model of a lithium battery in an ICD.

The model presented in Eq. (2) is Lipschitz continuous; therefore, the existence of the unique solution for this model is guaranteed, knowing the initial conditions. We solved this model using the Euler method and assumed that the Li/SVO-CFx initial depth of discharge is 5% (or the battery is almost fully charged with 95% state of charge), the battery initial temperature is equal to the patient body core temperature ($x_3(0) = 37°$), and the initial transient voltage of the battery is zero. Our results show that the cell's voltage dynamics are more affected by small changes in the constant low-energy continuous current drag than by high-energy pulsation. The results and discussions are presented in the next section.

TABLE II: Battery thermal model parameters

| Parameter | values, Unit |
|---|---|
| Q | 2000 [mAh] |
| h | 11 [$J/s.m^2 K$] |
| A | 3e-4 [$m^2$] |
| m | 0.03 [kg] |
| I | 25 [mu Amp] |
| $R_1$ | 0.11 [Ω] |
| $R_2$ | 2.28e-2 [Ω] |
| $C_1$ | 2.35e3 [Ω] |

## IV. RESULTS

In this paper the goal is to observe the battery terminal voltage, DOD and internal temperature dynamics over 5 to 8 years of battery drainage in an implanted ICD device by means of simulation on a lumped parameter equivalent circuit model coupled with a thermal model. ICDs generally last for 5 to 7 years and the end-of-service time of the device comes not at the stage of fully depleted battery capacity, but earlier when the terminal voltage of the battery drops below a certain level (at around 2.5 $V$, or after the knee in the voltage curve) [23], [24]. An average housekeeping constant current drainage of 25 $\mu A$ and a defibrillation current drainage of 3.0 A with an overestimated pulsation intervals of 3 months and pulse width of 10 S is applied to the model of a Li/SVO-CFx battery. These values are obtained from the work done by Root and Baliga [25], [26]. The initial battery voltage is 3.2 $V$.

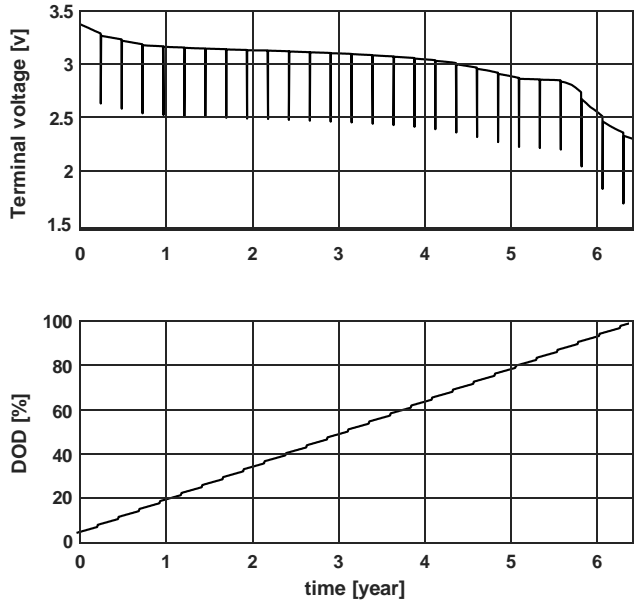

Fig. 4: Terminal voltage and depth of discharge (DOD) of a li/SVO-CFx battery under an average housekeeping and defibrillation load in an implantable cardioverter defibrillator (ICD).

Fig. (4) shows the simulation results for the terminal voltage and DOD. The voltage drop's magnitude is 0.7 $V$ for every defibrillation pulse and the battery transient voltage raises fast after the pulse because of the very small time constant of its dynamics which is $R_1 C_1 = 2.2e^{-3}$ S. The terminal voltage of the battery decreases gradually over the 6-year device lifetime, and numerous defibrillation pulses do not disturb the voltage dynamics. The DOD plot in Fig. (4) shows the linear loss of battery capacity over the device lifetime with a constant rate of 14.3% per year. Each defibrillation pulse discharges the battery by 2.2% per pulse. Fig. (5) also represents the battery internal temperature behavior during the device lifetime. The temperature barely goes beyond 37 $^\circ C$ for battery normal operation for device housekeeping; however, it raises to 37.42 $^\circ C$ with each defibrillation. This is because of the irreversible heat generation of the battery across its ohmic resistance. The thermal time-constant of the battery in this model is $\frac{mCp}{hA}$ which is very small and results in the fast return of the temperature to its equilibrium point. This simulation model captures the battery voltage and DOD dynamics during a normal battery drainage in an ICD. It predicts battery failure in an ICD device by means of accurate estimation of battery voltage. The battery will fail to deliver the required housekeeping or defibrillation energy if its voltage drops below 2 $V$. This model simulation does not capture external battery failure causes such as current leakage due to ICD's lead insulation failure and device malfunctions. It also does not capture internal battery failure phenomena such as battery internal short circuits or loss of active materials due to parasitic side reactions. However, the model can potentially be used in ICD battery current control to predict the battery voltage and DOD, so that premature device failure can be prevented. The model can also improve the long-term battery performance prediction for design purposes. We apply a range of discharge current magnitudes for both the defibrillation pulses and the housekeeping current to see their impact on the battery voltage and the device lifetime. The goal is to identify which one contributes more to battery depletion.

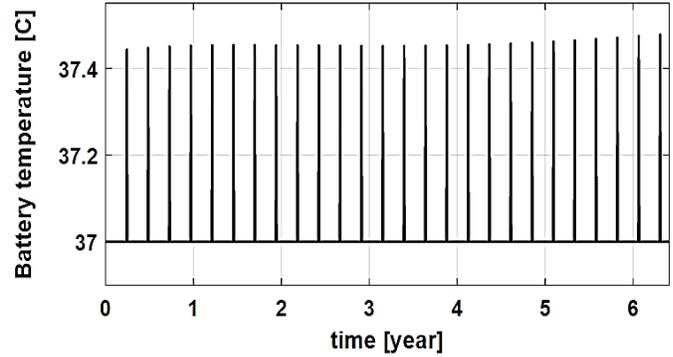

Fig. 5: Battery internal temperature trajectory for an average housekeeping and defibrillation load in an ICD.

In Fig. (6), we increased the defibrillation energy delivery from 3 $A$ to 8 $A$, in increments of 1 $A$. The solid line is for the normal 3 $A$ defibrillation, and it shows a lifetime of 7 years for the device (the knee of the time series voltage curve occurs after 7 years of battery use). The knee shifts to the left as the defibrillation pulse current increases; however, the magnitude of the shift is small. This figure shows that if the physician decides to double the defibrillation energy due to higher tissue impedance, the device lifetime will diminish by half a year. When the load is doubled (comparing the solid black line and the dashed line in Fig. (6)), the slope of the voltage plateau remains almost unchanged, especially during the beginning and midlife of the device.

We also studied the impact of doubling the housekeeping current on the ICD's battery performance. The addition of electrical components to an ICD, such as telemetry circuitry or measurement units takes more energy out of the bat- tery. The housekeeping current comprises all these battery usages for the ICD device to function. Here we increase this current from an average of 25 $\mu A$ to 65 $\mu A$ in increments of 10 $\mu A$ to see the impact on the battery terminal voltage and device lifetime. Fig. (7) shows the simulation results for this study. The solid black line shows results for a constant housekeeping current of 25 $\mu A$ and the dashed line shows results for a housekeeping current of 65 $\mu A$. The defibrillation current stays at 3 $A$ for all five simulations.

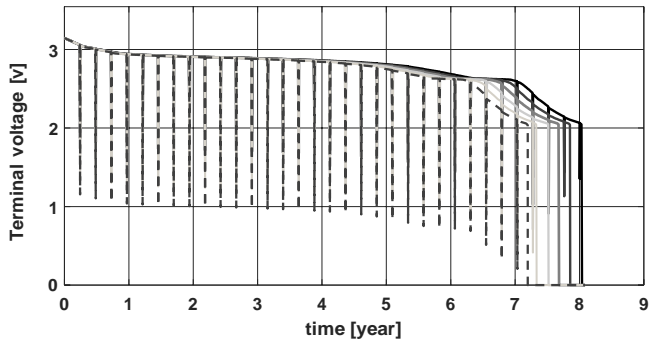

Fig. 6: Impact of raising the defibrillation discharge current on the ICD battery voltage. (from the solid black line to the dashed line, the defibrillation current increments by 1 $A$ from 3 $A$ to 8 $A$.)

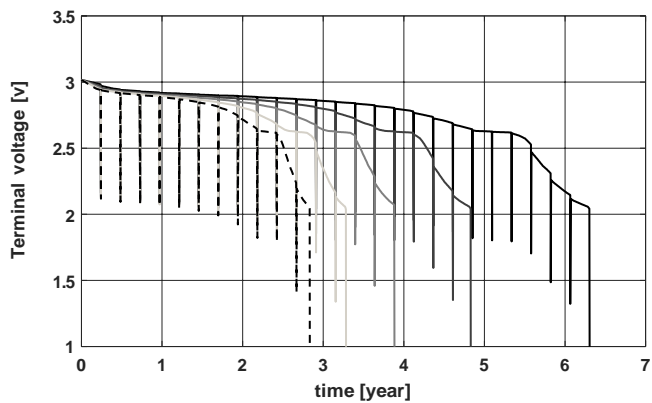

Fig. 7: Impact of raising the housekeeping discharge current on the ICD battery voltage. (From the solid black line to the dashed line, the defibrillation current increments by 10 $\mu A$ from 25 $\mu A$ to 65 $\mu A$)

Fig. (7) presents an interesting outcome: A linear increase in the constant discharge current of the ICD's battery in increments of 10$muA$ results in a nonlinear knee shift to the left. It means that an ICD device will last for about 5.5 years with an average housekeeping current of 25 $\mu A$ and it lasts 1.5 years less for the first 10 $\mu A$ increase in housekeeping current, 2 years less for the second increment of 10 $\mu A$, 2.5 years less for the third increment of 10 $\mu A$, 2.7 years less for the forth increment of 10 $\mu A$, and 3 years less for the last increment of 10 $\mu A$. This suggests that monitoring the impact of additional housekeeping current on the ICD's battery depletion is important to avoiding unexpected voltage drops and depletion. The simulation also shows that doubling the very small housekeeping discharge current from 25 $\mu A$ to 55 $\mu A$ reduces the battery's lifetime by almost half. It also demonstrates that at higher housekeeping currents, the rate of voltage drops increases as it can be seen in the dashed line compared to the solid black line voltage curves. Interestingly, in comparison with doubling the defibrillation discharge current in Fig. (6), this result shows the impressive effect of small housekeeping constant discharge current on battery performance and lifetime. Doubling the defibrillation current diminishes only 14% of the battery lifetime but doubling the housekeeping current reduces the battery life by 50%.

## V. Discussion

This paper presents a simulation analysis to obtain insights into the operation of Li/SVO-CFx batteries in ICDs for a range of defibrillation and housekeeping currents. We applied the load current to a coupled electrothermal model of the battery to simulate the battery terminal voltage, DOD and internal temperature over the average lifetime of the ICD device (5 to 8 years). Based on our analysis, the batteries deplete faster with increasing housekeeping current than with increasing defibrillation current. Additional defibrillation pulsation current by a factor of two does not change the rate of voltage drop across the battery terminals and does not diminish the battery lifetime by more than 14%. However, a factor of two increase in the housekeeping constant current depletes the battery two times faster. Our study results show that the internal temperature of the Li/SVO-CFx doesn't increase more than 0.1% over the entire battery lifetime in an ICD due to housekeeping current. The defibrillation current pulses, however, increases the internal battery temperature by 0.42 $^\circ C$. This sudden change in battery temperature does not last long and will get back to 37 $^\circ C$ quickly due to the very small time-constant of the battery temperature dynamics. Overall, our model and results show how simulations of ICD battery dynamics can aid in preventing sudden battery voltage drop off and to better predict the lifetime of the device.

## Acknowledgements

This work was supported in part by an intramural FDA Chief Scientist Challenge Grant and an appointment to the Research Participation Program at the Center for Devices and Radiological Health administered by the Oak Ridge Institute for Science and Education through an interagency agreement between the U.S. Department of Energy and the U.S. Food and Drug Administration. The authors gratefully acknowledge this support. Any opinions, findings, and conclusions or recommendations expressed in this material are those of the authors and do not necessarily reflect the views of the Food and Drug Administration. Moreover, the mention of commercial products, their sources, or their use in connection with material reported herein is not to be construed as either an actual or implied endorsement of such products by the Department of Health and Human Services.